\documentclass[aps,english,prb,reprint,superscriptaddress, citeautoscript,cite]{revtex4-1}

\usepackage{graphicx}
\usepackage{amssymb}
\usepackage{babel}
\usepackage{color}
\usepackage{multirow}

\begin{document}

\title{Modification of electronic, magnetic structure and topological phase of bismuthene by point defects}

\author{Yelda Kadioglu}
\affiliation{Department of Physics, Adnan Menderes University, Ayd{\i}n 09010, Turkey.}

\author{Sevket Berkay Kilic}
\affiliation{Department of Physics, Bilkent University, Ankara 06800, Turkey}

\author{Salih Demirci}
\affiliation{UNAM - Institute of Materials Science and Nanotechnology, Bilkent University, 06800 Ankara, Turkey}
\affiliation{Department of Physics, K\i r\i kkale University, K\i r\i kkale 71450, Turkey}

\author{O. \"Uzengi Akt\"urk}
\affiliation{Department of Electrical and Electronic Engineering, Adnan Menderes University,
09100 Ayd{\i}n, Turkey}
\affiliation{Nanotechnology Application and Research Center, Adnan Menderes University, Ayd{\i}n 09010, Turkey}

\author{Ethem Akt\"{u}rk}\email{ethem.akturk@adu.edu.tr}
\affiliation{Department of Physics, Adnan Menderes University, Ayd{\i}n 09010, Turkey.}
\affiliation{Nanotechnology Application and Research Center, Adnan Menderes University, Ayd{\i}n 09010, Turkey}

\author{Salim Ciraci}\email{ciraci@fen.bilkent.edu.tr}
\affiliation{Department of Physics, Bilkent University, Ankara 06800, Turkey}

\date{\today}

\begin{abstract}
This paper reveals how electronic, magnetic structure and topological phase of 2D, single-layer structures of bismuth are modified by point defects. We first showed that free standing, single-layer, hexagonal structure of bismuth, named as h-bismuthene exhibits non-trivial band topology. We then investigated interactions between single foreign adatoms and bismuthene structures, which comprise stability, bonding, electronic and magnetic structures. Localized states in diverse location of the band gap and resonant states in band continua of bismuthene are induced upon the adsorption of different adatoms, which modify electronic and magnetic properties. Specific adatoms result in reconstruction around the adsorption site. Single and divacancies can form readily in bismuthene structures and remain stable at high temperatures. Through rebondings Stone-Whales type defects are constructed by divacancies, which transform into a large hole at high temperature. Like adsorbed adatoms, vacancies induce also localized gap states, which can be eliminated through rebondings in divacancies. We also showed that not only optical and magnetic properties, but also topological features of pristine h-bismuthene can be modified by point defects. Modification of topological features depends on the energies of localized states and also on the strength of coupling between point defects.
\end{abstract}

\maketitle

\section{Introduction}
Three-dimensional (3D) quasi layered bismuth crystal being the heaviest element in group V column or pnictogens have attracted the interest of many researchers owing to their exceptional features.\cite{ast,koroteev,tichovolsky,perfetti} It has the highest resistivity and Hall coefficient of all metals and highly anisotropic Fermi surface.\cite{gallo,kim2013,shim,ast2001} Narrow fundamental band gap with seizable spin-orbit coupling (SOC) makes 3D and its surfaces crucial topologically. In this respect, the topological behavior of 3D Bi and compounds, as well as its thin films grown on specific substrates have been studied actively.\cite{nagao,blugel2,yaginuma,wada,zlui,fyang,lchen,lcheng} Two-dimensional Bi films of buckled honeycomb and black phosphorus-like washboard structure have Rashba-type splitting of the surface states due to strong SOC. This leads to a large variation of electronic properties ranging from narrow band gap semiconducting to semimetallic and metallic states.\cite{blugel2,yaginuma} All thin films of buckled honeycomb structure have been found to be topological insulator.\cite{zlui} Furthermore, topological insulator property has remained robust under the applied electric field and strain.\cite{lchen} Not only bare surfaces of Bi, but also the surfaces covered with H have been studied to reveal the effect of adsorbed hydrogen on the topological character of the material.\cite{freitas}

As the synthesis of phosphorene, i.e. 2D, single-layer (SL) structures of phosphorus, and devices fabricated therefrom gain importance, similar structures have been predicted for other group V elements such as  arsenene \cite{kamal}, antimonene \cite{antimon}, nitrogene.\cite{nitro} Most recently, based on extensive dynamical and thermal stability analysis, stable, free-standing, 2D SL phases of bismuth, namely buckled honeycomb or hexagonal (h-Bi), symmetric washboard (w-Bi) and asymmetric washboard structures (aw-Bi), which are identified as \textit{bismuthene}, were unveiled, with aw-Bi having total energy lower than w-Bi.\cite{ethem} Once the stability of free standing h-Bi and aw-Bi has been demonstrated, whether they are topologically trivial remained to be examined.

The electronic structures of h-Bi and aw-Bi can be modified when they are placed on a suitable substrate with minute substrate overlayer interaction, or when they form bilayers and multilayers whereby the fundamental band gap can be tuned with the numbers of layers. Another efficient way of modifying electronic and magnetic properties of single and multilayers can be the creation of point defects, such as adatom adsorption (or surface doping) and single and divacancy creation. At low coverage of point defects, the coupling between adjacent defects is minute and give rise to localized states in the band gap and local magnetic moments. This way, SL or multilayer (ML) bismuthene attains localized gap states. Additionally, trivial-to-topological transitions or vice versa can also occur.

In this paper, we investigated the effect of a single, isolated point defects on the electronic and magnetic properties, as well as on topological phases of  2D SL h-Bi and aw-Bi. As point defects, we considered the adsorption of selected foreign adatoms, H, C, O, Si, P, Ge, As, Se, Sb, Pb, Sn, Te, as well as single and divacancy formation within supercell geometry. Major effort is devoted to find the equilibrium adsorption geometry namely optimized structures of host atoms and adatoms with lowest total energy. Similarly, we calculate also the optimized atomic structure at the close proximity of vacancy and divacancy and corresponding formation energies. Additionally, we attempted to answer the crucial question of how the topological phase of pristine, free standing bismuthene is modified by selected point defects.

Important findings of our study are summarized as follows: (i) While free standing, SL h-Bi structure is topologically trivial, aw-Bi is not. (ii) The adatoms mentioned above form strong chemical bonds with binding energies ranging from $\sim$ 1 eV to $\sim$ 3.5 eV. Except local deformations at the adsorption site, SL bismuthene structures remain stable. (iii) Localized states, the energies of which are adatom specific, can occur in the band gap. The electronic properties of bismuthene are affected by these gap states. Adsorbed P and Sb can attribute magnetization to bismuthene. (iv) Single vacancies can form readily in bismuthene and give rise to localized and resonant states, whereby the band gap is further reduced. Owing to the rebonding among Bi host atoms surrounding the divacancy, the dangling bonds at the defect site are saturated and hence the localized states in the gap are eliminated. (v) A divacancy in h-Bi transform into Stone-Whales type defect and to large holes at high temperature. (vi) Depending on the type and coupling between the point defects, topologically non-trivial phase of the bare bismuthene can be destroyed.

\section{Atomistic Model and Computational Details}
Here we consider the low vacancy concentration and low coverage of adatoms in order to minimize couplings between adjacent defects and to reveal the effect of single defects on the physical properties. This situation is mimicked within periodic boundary conditions and supercell geometry using $(5 \times 5)$ supercells, each one including a single point defect. Within this supercell geometry, the distance of $\sim$22 \AA between point defects in the adjacent supercells is assured.

\begin{figure*}
\includegraphics[scale=0.8]{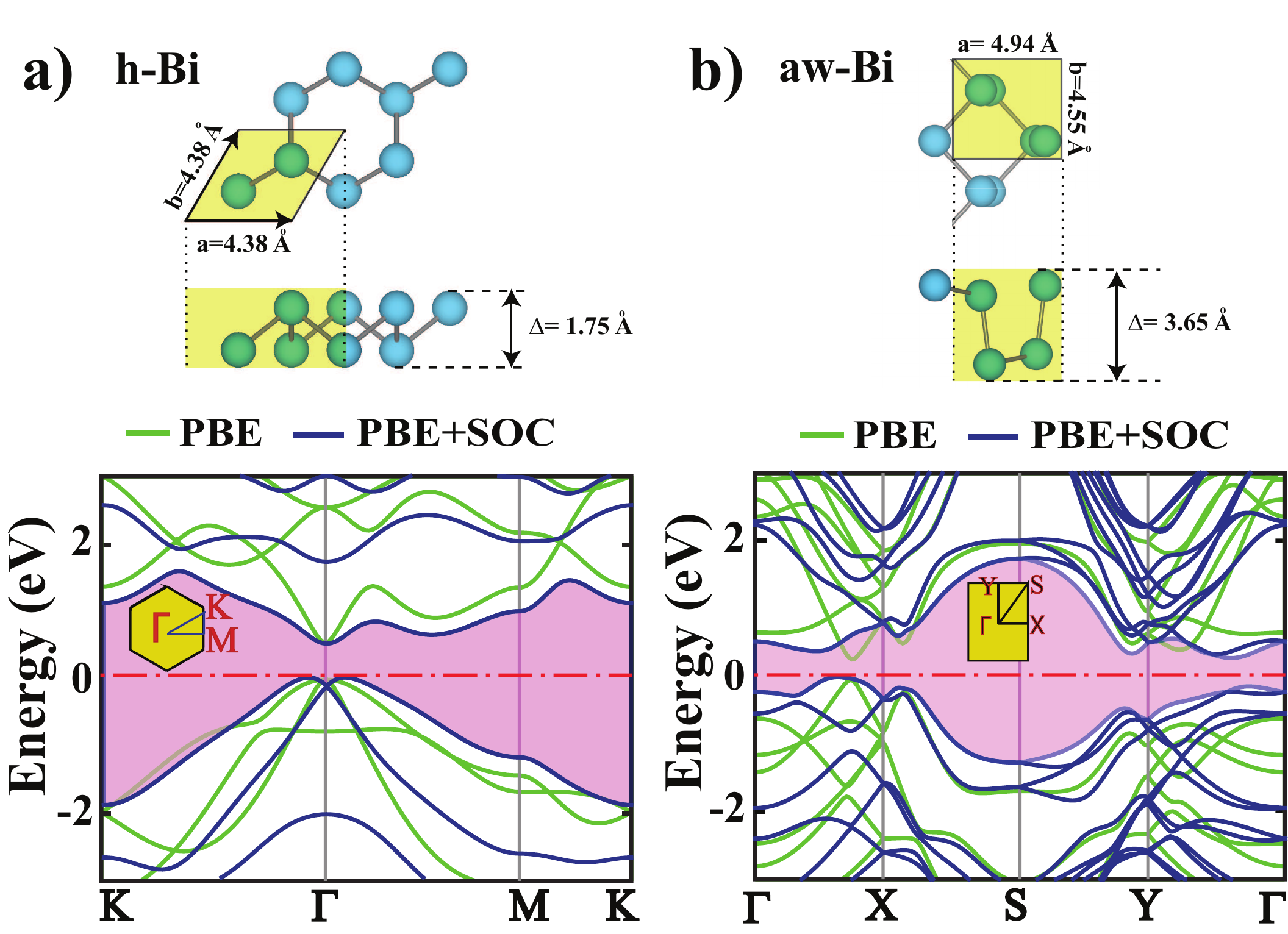}
\caption{(a) Top and side views of the unit cell of SL h-Bi with optimized lattice constants and buckling $\Delta$. Electronic energy band structures of bare SL h-Bi structure calculated using PBE, PBE+SOC. Zero of band energy is taken at the top of the valence band. (b) Same for SL aw-Bi.}
\label{fig1}
\end{figure*}

Our theoretical analysis and predictions are obtained by first-principles plane-wave calculations based on spin-polarized density functional theory (DFT) within generalized gradient approximation (GGA). The Perdew-Burke-Ernzerhof (PBE) functionals were used for the exchange-correlation potential \cite{pbe} and the PAW pseudo potentials were adopted \cite{paw1,paw2}. Plane wave basis sets with a kinetic energy cutoff of 500 eV was used. Monkhorst-Pack \cite{monk} mesh of 4x4x1 were employed for the Brillouin zone (BZ) integration. D2-Grimme correction (DFT-D2) \cite{Grimme2006} have been taken into account for London dispersion corrections. Spin orbit coupling (SOC) is included in all calculations.
Atomic positions were optimized using the conjugate gradient method until Hellmann-Feynman forces acting on each atom less than 0.02 eV/\AA. The maximum pressure in the unit cell was reduced to less than 1 kbar. The energy convergence criterion of the electronic self-consistency was chosen as $10^{-5}$ eV between two successive iterations. Gaussian type Fermi-level smearing method is used with a smearing width of 0.01 eV. Numerical calculations were performed using VASP.\cite{vasp1,vasp2}

Binding energies of adatoms (A), which are crucial for the strength of interaction between bismuthene and adatom, are calculated using the expression, $E_b = E_T[h(aw)-Bi]+E_T[A]-E_T[h(aw)-Bi + A]$ in terms of the total energies (per cell) of bare bismuthene (specified as h(aw)-Bi), $E_T[h(aw)-Bi]$; free adatom, $E_T[A]$; adatom adsorbed bismuthene, $E_T[h(aw)-Bi + A]$. Here the positive values of $E_b$ indicates a binding structure. The formation of single vacancy is calculated using the expression,\cite{Gillan} $E_v=E_T[h(aw)-Bi + V]-\frac{m-1}{m} E_{T}[h(aw)-Bi]$ in terms of he total energies of bismuthene including one single vacancy in the supercell and bare bismuthene having $m$ Bi atoms in the supercell.

To test whether bare and defected free standing SL bismuthene are topologically trivial or non-trivial we carried out calculations with Wannier90 package and Z$_2$-Pack\cite{soluyanov,gresch}, using ab-initio VASP output. Z$_{2}$ invariant is calculated by tracking the time evolution of Hybrid Wannier Charge Centers (WCC) on the surface where the momentum, $k_{y}$ in half BZ is mapped to a time interval (0,$T/2$), and $k_{x}$ is used to construct the maximally localized Wannier Functions.

\section{Pristine SL Bismuthene Phases}
The energetics and electronic structures of bare bismuthene phases will serve as reference to unveil the modifications of its electronic structure with point defect. The lattice constants of 2D hexagonal lattice of SL h-Bi is 4.38 \AA~ with buckling parameter $\Delta$=1.75 \AA, and the cohesive energy $E_c$=1.95 eV/atom and the formation energy relative to 3D Bi crystal is $E_f$=-0.13 eV/atom.\cite{ethem,hirahara} The electronic structure calculated with PBE+SOC has indirect band gap of 0.51 eV in Fig.~\ref{fig1} (a). Single-layer aw-Bi phase is slightly more energetic\cite{ethem} with $E_c$=1.97 eV and $E_f$=-0.11 eV; the rectangular lattice constants of the optimized structure are 4.94 \AA~ and 4.55 \AA. The fundamental band gap is indirect and is calculated to be 0.28 eV in Fig.~\ref{fig1} (b). Apparently, both phases of SL bismuthene structures are narrow band gap semiconductors. More details about the phonon dispersion, elastic constants, electronic structure and their bilayers can be obtained from Ref.[\cite{ethem}] and references therein.

\begin{figure*}
\includegraphics[scale=0.5]{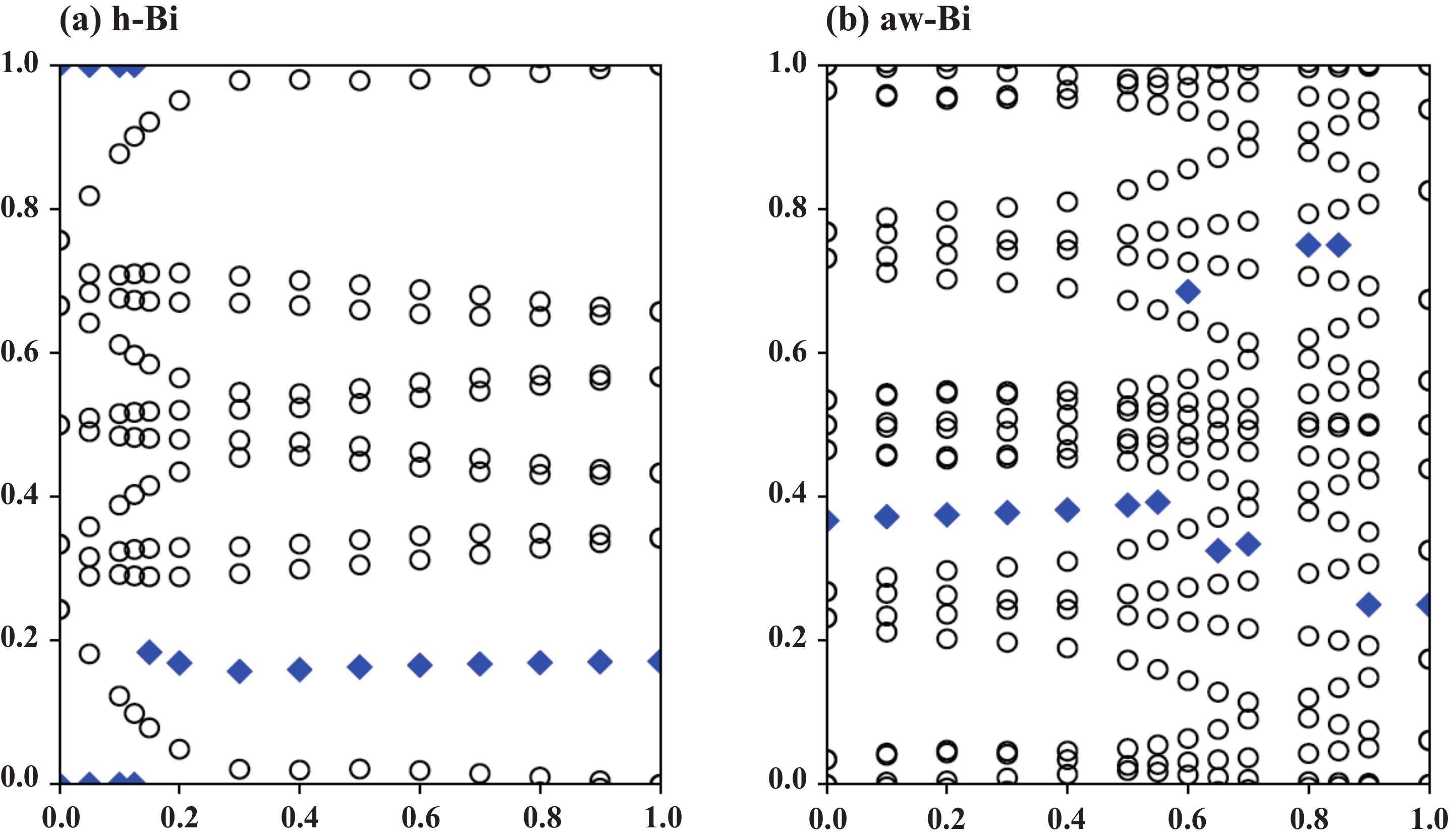}
\caption{Time evolution of Wannier Charge Center (WCC)s at the $k_z=0$ surface for (a) non-trivial free standing, SL h-Bi, and (b) trivial free standing, aw-Bi. The circles represent WCCs and the diamonds represent the midpoint of two closest neighboring WCCs\cite{soluyanov}. $x$- and $y$-axis axis represents the time interval and $y$-axis represent the time and position of the WCC, respectively. For h-Bi, the midpoint crosses WCCs odd number of (1) times, giving Z$_2$ invariant equal to 1 and for aw-Bi, midpoint crosses the WCCs even number of (34) times, resulting with the Z$_2$ invariant being 0.}
\label{fig2}
\end{figure*}

\subsection{Band topology of SL bismuthene}
Two dimensional topological insulators has been an active field of research due to promising spintronic applications. After graphene was predicted to possess the non-trivial topological  phase\cite{kane}, other 2D honeycomb materials such as silicene, were shown to be topological insulators\cite{ezawa}. Bi has SOC significant enough to induce a band inversion. 3D bulk Bi, its surfaces, as well as its thin films have been shown to posses the non-trivial phase under special conditions.\cite{nagao,blugel2,yaginuma,wada,zlui,fyang,lchen,lcheng} Also, h-Bi placed on Si(0001) substrate was shown to posses topologically non-trival phase.\cite{hsu}

To reveal the effect of point defects on the topological behavior, we first considered pristine, free-standing, SL h-Bi and aw-Bi structures. Using ab-inito calculations, we calculated Z$_2$ invariant for free standing SL h-Bi, and found that it is a topological insulator with Z$_2$ invariant equal to 1. Then we calculated the Z$_2$ invariant of free standing, SL aw-Bi structure, and found that Z$_2$=0. This means that aw-Bi has a trivial topological order. (See Fig.~\ref{fig2}). This shows that despite aw-Bi and h-Bi share the same host atom and have similar dimensionality, they have different band topologies. However, that aw-Bi can attain topologically non-trivial phase under selected electric field, strain or compound formation is possibility worth investigating under a different context.

\begin{figure*}
\includegraphics[scale=0.8]{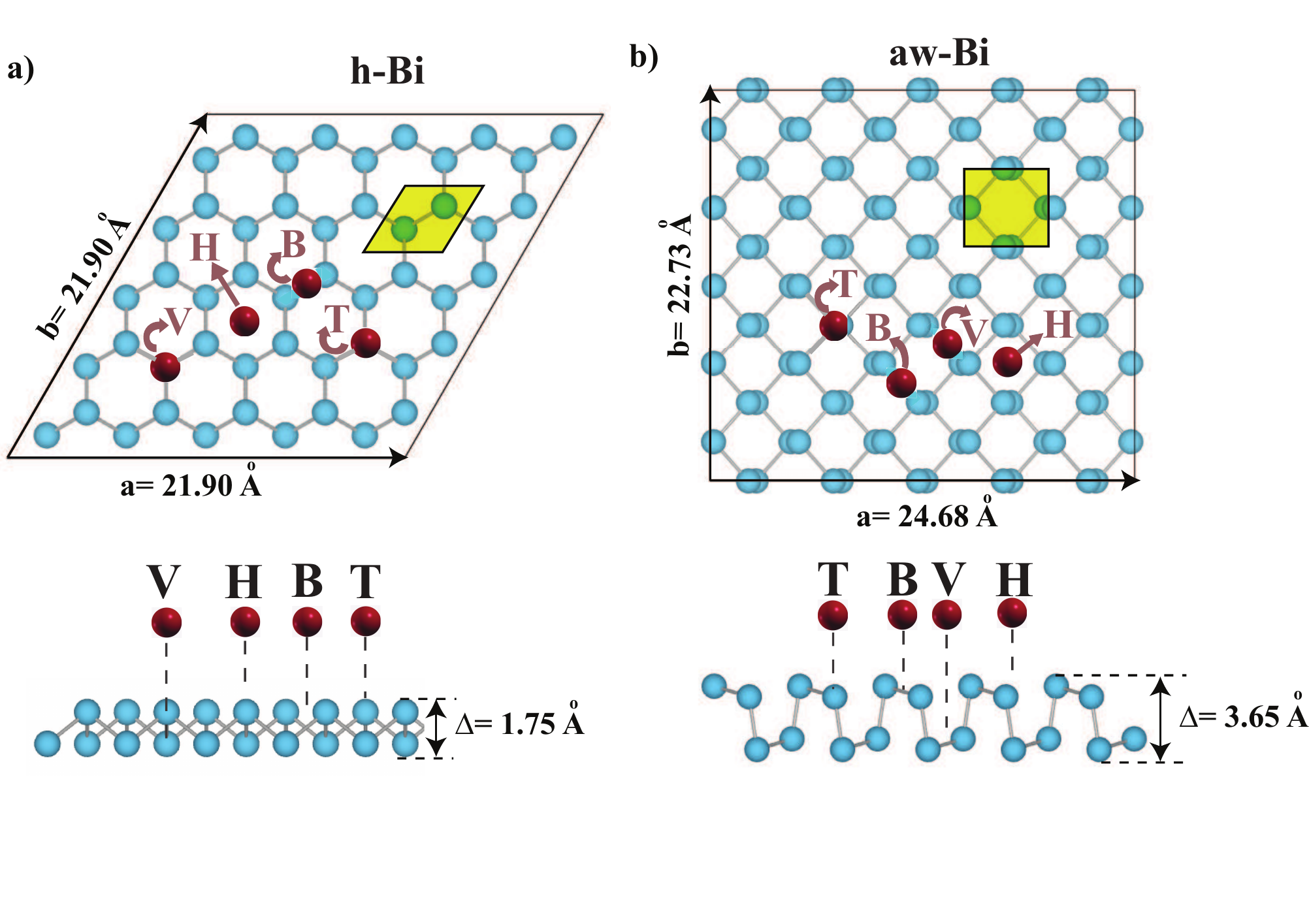}
\caption{(a) Description of possible adsorption sites, T,H,B,V, on the $(5 \times 5)$ supercells of SL h-Bi in top and side view. Primitive unit cells are shaded. Lattice constants of the supercell and $\Delta$ buckling parameters are also shown. (b) Same for SL aw-Bi.}
\label{fig3}
\end{figure*}

\section{Adatom Adsorption}
Here we investigate the interaction of adatoms, H, C, O, Si, P, Ge, As, Se, Sb, Pb, Sn and Te, with bismuthene phases. Most of these adatoms share the same column as Bi. Group IV atoms, which makes important SL structures, are also included in this study. Interaction of H and O, is important because of possible hydrogenation and oxidation processes in SL and ML bismuthene phases. Adsorbed Se and Te are expected to offer features interesting in opto-electronic applications.

Possible adsorption sites on the supercell of SL h-Bi and aw-Bi are described in Fig.~\ref{fig3}. These are on top of Bi atom, i.e. top site (T); on top of the center of hexagon, i.e. hollow site (H); on top of low-buckled Bi atom, i.e. valley site (V); on the center of Bi-Bi bond, i.e. bridge site (B). For a given adatom, we first determine the configuration of equilibrium adsorption site, namely a particular site with well-defined height $z$, and lateral position $(x,y)$ yielding the highest binding energy. For each adatom, the equilibrium site and corresponding binding energy is calculated after a comprehensive optimization processes comprising the positions of adatom and all the host Bi atoms in the supercell. In Table~\ref{tableI} and Table~\ref{tableII} equilibrium adsorption site, binding energy, magnetic moment and other data calculated for adatom adsorbed to SL h-Bi and aw-Bi are presented.

\begin{table}
\caption{ Calculated values for the adatom adsorbed to SL h-Bi. Optimized equilibrium site; binding energy, $E_{b}$ with/without vdW correction; the height (distance) of the adatom from the original, high-lying Bi atomic plane of h-Bi, $h$; the minimum distance between the adatom and host Bi atom of bismuthene, $d_{A-Bi}$; the magnitude of the local magnetic moment $\mu$.}
\begin{tabular}{ccccccc}
Adatom (A) & Site & $E_{b}$ (eV) & $h$(\AA) &$d_{A-Bi}$(\AA) & $\mu$ ($\mu_B$) \tabularnewline
\hline
H         & H &  1.33/ 1.21  & 1.10 & 1.91 & 0 \tabularnewline
C         & V &  3.70/ 3.48  &-0.43 & 2.30 & 0 \tabularnewline
O         & Br &  4.08/ 3.54  & 0.30 & 2.17 & 0   \tabularnewline
Si        & H-V &  3.17/ 2.33  &-0.76 & 2.61 & 0  \tabularnewline
P         & V &  2.23/ 2.00  & 1.12 & 2.62 & $\sim$1 \tabularnewline
Ge	  & V &  2.11/ 1.90  & 1.17 & 2.96 & 0 \tabularnewline
As	  & H-V &  2.07/ 1.82  & 1.27 & 2.71 & $\sim$1 \tabularnewline
Se        & H &  2.44/ 2.22  & 1.79 & 2.92 & 0   \tabularnewline
Sb        & H-T &  2.06/ 1.14  & 1.49 & 2.90 & $\sim$1 \tabularnewline
Pb	  & V &  1.13/ 0.81  & 1.91 & 3.29 & 0 \tabularnewline
Sn        & H-T &  1.88/ 1.54  & 1.40 & 3.05 & 0 \tabularnewline
Te        & T-Br &  1.91/ 1.57  & 1.47 & 2.79 & 0 \tabularnewline
\hline \hline
\label{tableI}
\end{tabular}
\end{table}

\begin{table}
\caption{ Calculated values for the adatom adsorbed to aw-Bi. Optimized equilibrium site; binding energy $E_{b}$ with/without vdW correction; the height (distance) of the adatom from the original, high-lying Bi atomic plane of aw-Bi, $h$; the minimum distance between the adatom and host Bi atom of bismuthene, $d_{A-Bi}$; the magnitude of local magnetic moment $\mu$.}
\begin{tabular}{ccccccc}
Adatom (A) & Site & $E_{b}$ (eV) & $h$(\AA) &$d_{A-Bi}$(\AA) & $\mu $ ($\mu_B$)      \tabularnewline
\hline
H          & V &   1.31/ 1.24 & 0.16 & 2.00 & 0 \tabularnewline
C          & H &   3.41/ 3.22 & 0.20 & 2.28 & 0  \tabularnewline
O          & Br &   3.91/ 3.75 & 0.62 & 2.15 & 0  \tabularnewline
Si         & H &   2.92/ 2.67 & 0.97 & 2.73 & 0  \tabularnewline
P          & H &   2.41/ 2.00 & 0.72 & 2.65 & $\sim$1 \tabularnewline
Ge	   & H &   2.66/ 2.19 & 1.12 & 2.80 & 0   \tabularnewline
As	   & H &   2.21/ 1.90 & 0.85 & 2.75 & $\sim$1 \tabularnewline
Se         & H-T &   2.57/ 2.34 & 1.32 & 2.66 & 0 \tabularnewline
Sb         & H &   1.96/ 1.60 & 1.12 & 2.93 & $\sim$1 \tabularnewline
Pb	   & H-T &   1.43/ 1.16 & 1.57 & 3.08 &  0 \tabularnewline
Sn         & H &   2.20/ 2.01 & 1.29 & 3.00 & 0  \tabularnewline
Te         & Br &   1.99/ 1.65 & 1.81 & 2.94 & 0  \tabularnewline
\hline \hline
\label{tableII}
\end{tabular}
\end{table}

\begin{figure*}
\includegraphics[scale=0.8]{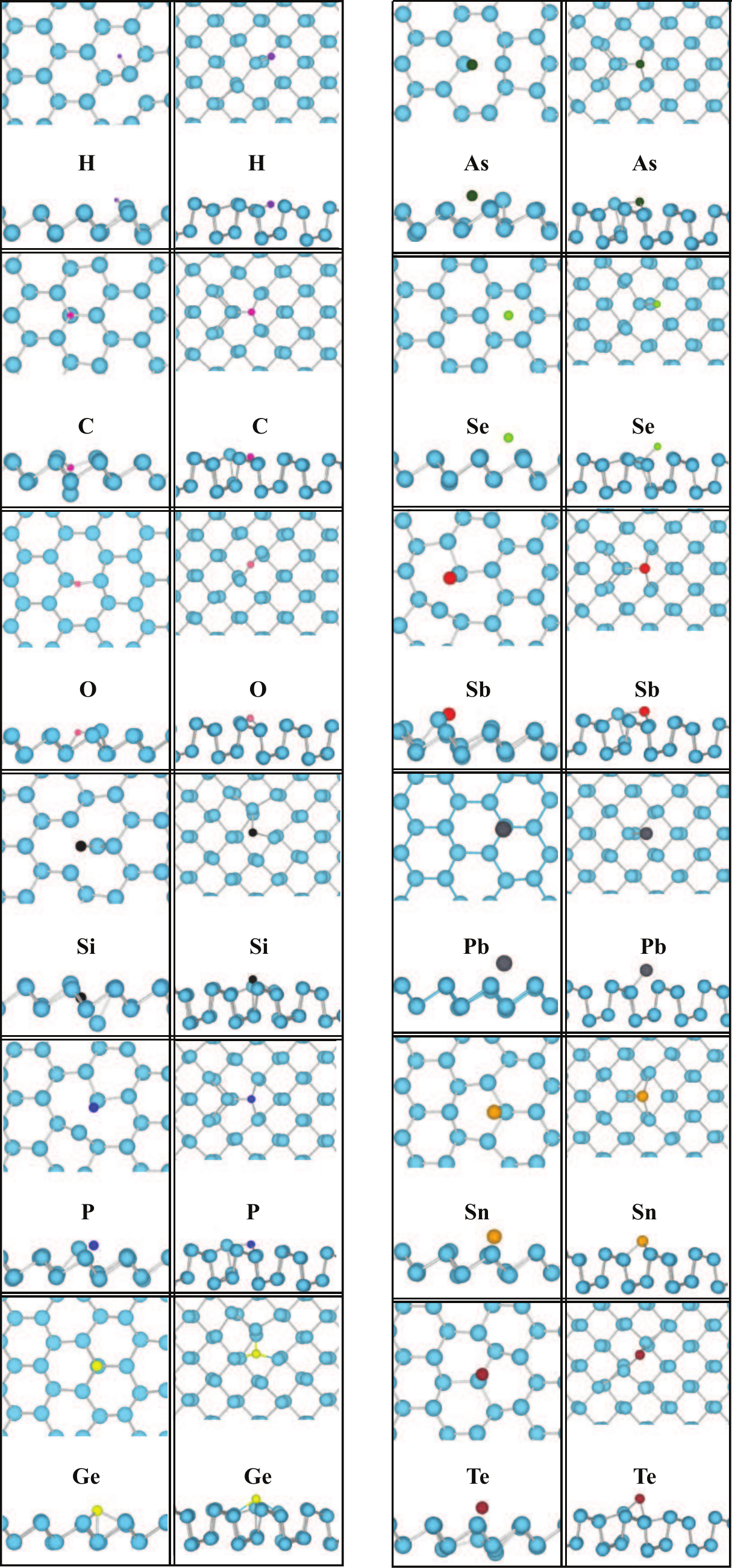}
\caption{Top and side views of the equilibrium atomic configuration of adatoms adsorbed to SL h-Bi (first and third columns) and SL aw-Bi (second and fourth columns). Host Bi atoms are shown by large, turquoise balls.}
\label{fig4}
\end{figure*}

The binding energies are high and are dominated by chemical bonding through charge exchange between the adatom and nearest host Bi atoms. The contribution of vdW interaction to the binding energies is generally small and ranges between 100 meV and 500 meV; but it is significant in the bonding of Si and Sb to h-Bi. Binding energies range from 1.1 ev to 4.1 eV. Carbon adatom has highest, Pb second highest binding energy. Adsorbed adatoms cause to local deformations of bismuthene. The equilibrium configurations of adatoms are shown in Fig.~\ref{fig4}.

Since the present study is considering the weak adatom-adatom coupling and attempting to mimic the chemisorption of a single, isolated adatom, the study of high-coverage limit or decoration of adatoms according to a given pattern causing significant adatom-adatom couplings is beyond the scope of this paper. Here, the effect of adatoms on the electronic properties of bare bismuthene phases are investigated by comparing (i) the total density of states (TDOS) of the adatom adsorbed bismuthene; (ii) density of states (DOS) projected to the adatom at equilibrium position; (iii) DOS projected to a Bi host atom farthest to the adatom. The latter DOS determines the electronic energy structure of the bare and extended SL bismuthene phases relative to the common Fermi level obtained from adsorbed adatom+SL bismuthene phase. In this respect, energies of adatom induced localized gap states and resonant states relative to the DOS of the bare and extended bismuthene can be revealed with a reasonable accuracy.

\begin{figure*}
\includegraphics[scale=0.8]{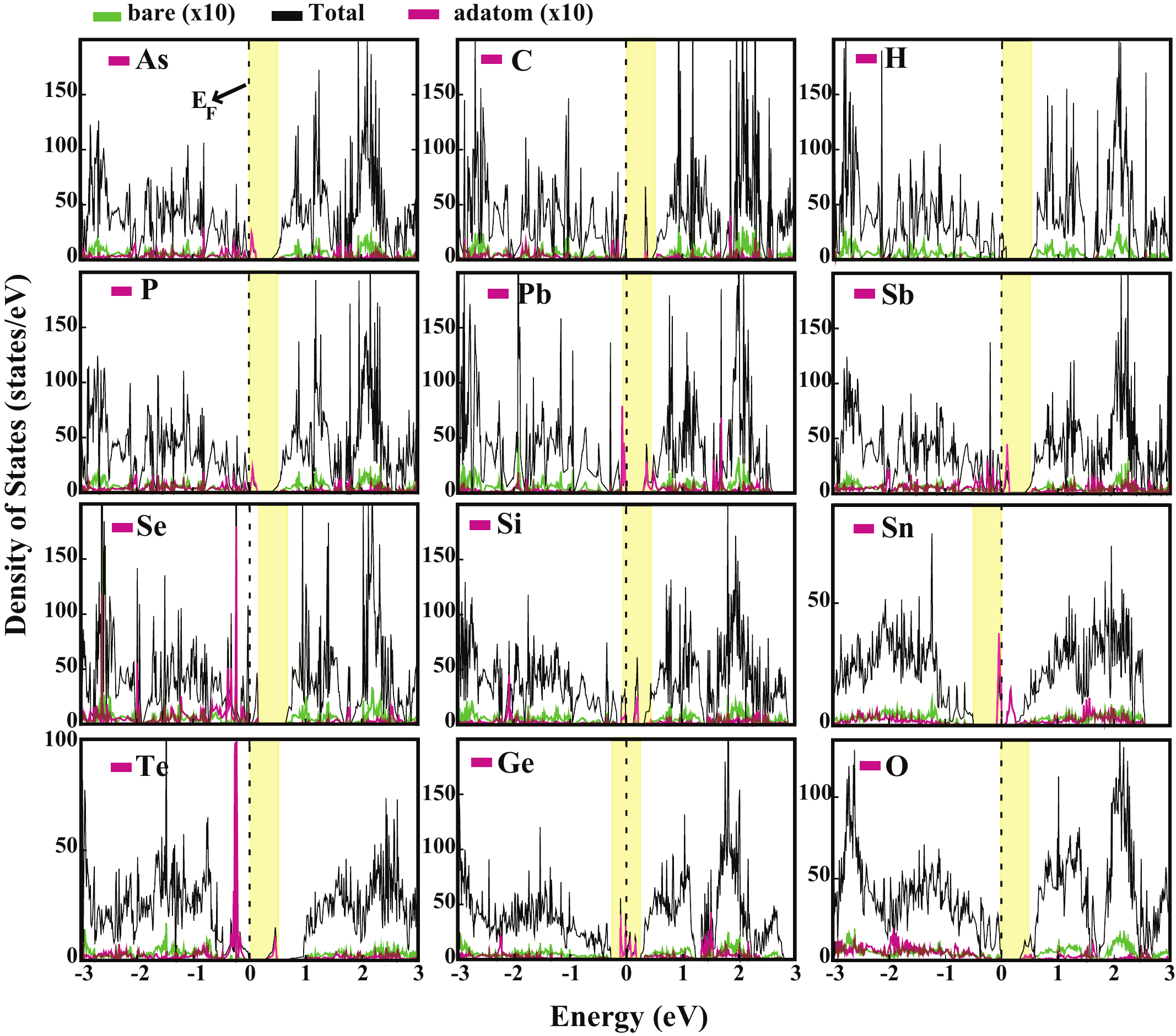}
\caption{Total density of states (TDOS) calculated for a single adatom adsorbed to each $(5 \times 5)$ supercell of h-Bi at the corresponding equilibrium site is shown by thin black lines. The density of states of the "bare'' extended h-Bi substrate is shown by light blue tone; its band gap is shaded by yellow zone. Pink lines indicate TDOS projected to the adatom. The zero of the energy is set at the common Fermi level shown by dashed vertical line.}
\label{fig5}
\end{figure*}

In Fig.~\ref{fig5}, we analyze total and projected densities of states of various adatoms adsorbed to h-Bi corresponding to the equilibrium configurations presented in Fig.~\ref{fig4}. The resonant states in the band continua, and localized gap states shown in detail modify the electronic structure of SL bare h-Bi. Here some crucial features are emphasized: Group V adatoms P, As, Sb give rise to localized gap states close to the top of the valence band. Adatoms, C, O, Pb, and Te, have localized gap states close to the conduction band edge. Additionally C, Se and Te have resonant states at the top of the valence band; but the resonant states of Sn is near the bottom of the conduction band. Localized gap states of Si and Ge occur at the midgap. Calculated common Fermi level determines whether a localized gap state is filled or empty.
\begin{figure*}
\includegraphics[scale=0.8]{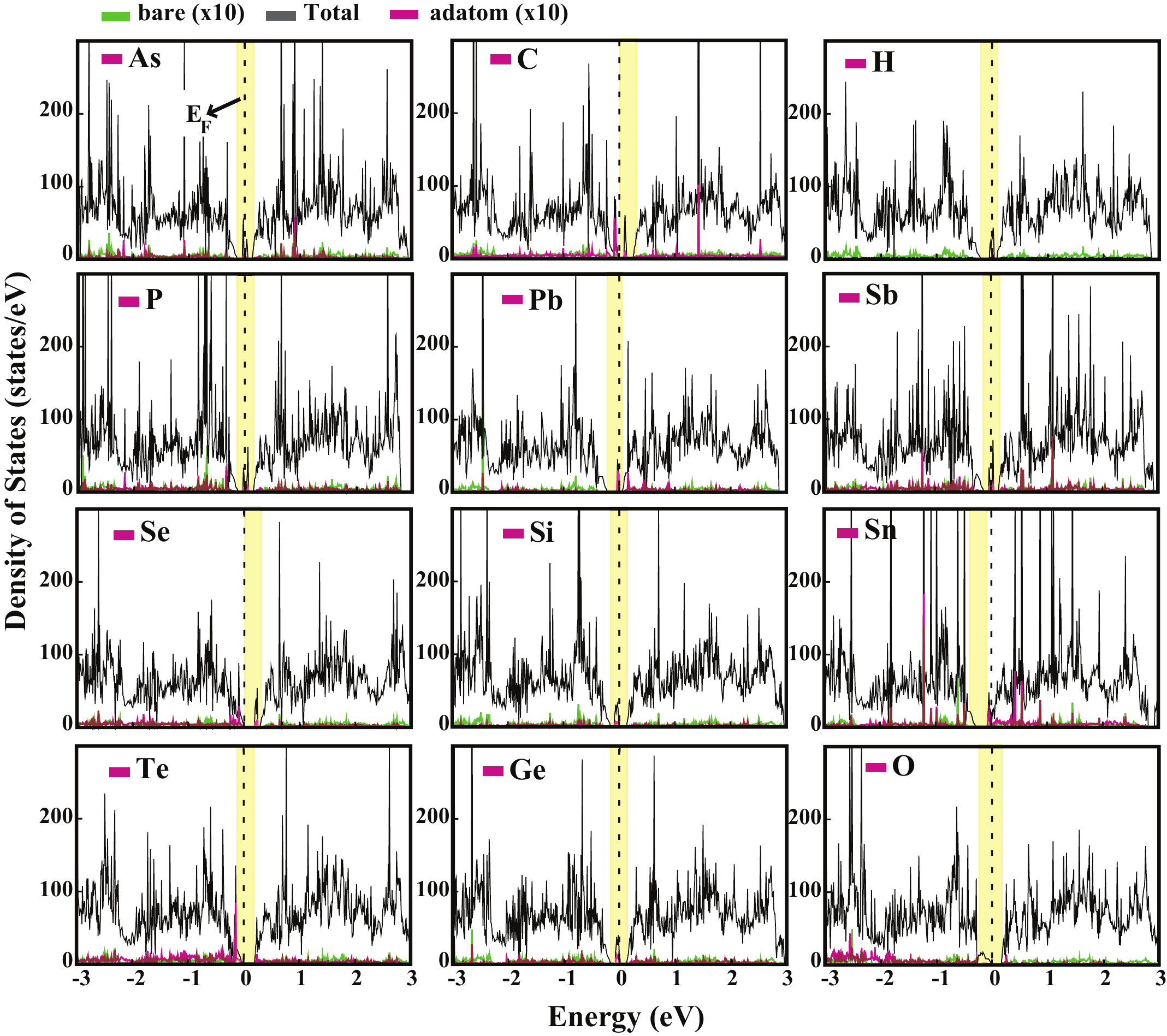}
\caption{Same as Fig.~\ref{fig5} for a single adatom adsorbed to each $(5 \times 5)$ supercell of aw-Bi.}
\label{fig6}
\end{figure*}

As shown in Fig.~\ref{fig6}, bare SL aw-Bi has relatively smaller PBE+SOC band gap. Adsorbed adatoms give rise to resonant and localized states similar to that of SL h-Bi, except some energy shifts. Here, the adsorption of H, As, Sb, P, Si and Ge, result in the states localized at the mid gap of the extended 2D SL aw-Bi. O, Se and Te give rise to states localizes near the valence and conduction band edges in the gap. Adsorbed C atom has localized states near the top of the valence band and at the midgap; adsorbed Pb  at the mid gap and the bottom of the conduction band. In addition, Sn has resonances in the conduction band continua, whereas strong resonances of Te occur at the top of the valence band.

To investigate the magnetic state of all adsorbed adatoms we performed a series of total energy calculations with SOC; each calculation had a different preset magnetic moment. For Group V adatoms we had to use larger energy cut-off to achieve convergence. The spin-polarized magnetic state and the permanent magnetic moment thereof were determined from the optimized structure, that resulted in the minimum total energy. Among all adatoms treated in this study, only Group V (pnictogens) adatoms P, As and SB adsorbed to h-Bi and aw-Bi had spin polarized ground state and attained permanent magnetic moments of $\sim$1 $\mu_B$ (see the last column in Table~\ref{tableI} and Table~\ref{tableII}). When free, Group V atoms, namely P, As, Sb have magnetic moments of 3.0 $\mu_B$. However, when they are adsorbed to h-Bi and aw-Bi their magnetic moments decreased because of the bond formation between unpaired orbitals of pnictogens with the orbitals of the substrate. Notably, projected densities of states of adatoms P, As and Sb adsorbed to h-Bi and aw-Bi presented in Fig.~\ref{fig5} and in Fig.~\ref{fig6} are not spin-polarized, since the orbitals or the density of states cannot be divided anymore into purely spin-up and spin-down contributions if SOC is included.

We note that at high-coverage limit or patterned decoration sharp localized states are broadened or flat impurity bands may form in the fundamental band gap. Coupling between adatoms and hence the size of the supercell determines the widths of impurity bands as well as the binding energies. Hence, the coverage and pattern of adsorption provides additional parameters to control the electronic structure. In summary, bare SL h-Bi and aw-Bi, attain new electronic states through the adsorption of adatoms, which may be suitable for 2D electronics. Functionalization of these nanostructures by the adsorption of adatoms is also important for their bilayers and multilayers.

\section{Vacancy and Divacancy Formation}
The single and divacancies in h-Bi are treated using $(5 \times 5)$ super cells and their formation energies are calculated to be 1.10 eV and 1.44 eV, respectively. The formation energy of divacancy is smaller than twice the formation energy of single vacancy, since number of bond broken in the former is smaller than the twice number of bonds in the latter. In Fig.~\ref{fig7} {a) the optimized atomic structures and corresponding electronic structure are presented. After the structure optimization, three Bi atoms surrounding the single vacancy slightly collapses towards the center. The stability of the single vacancy is assured by \textit{ab-initio} molecular dynamics calculations performed at T=500 K. Snapshots taken in the course of simulation shows that even if Bi atoms surrounding the vacancy are displaced from their equilibrium positions, overall character of the honeycomb structure is remained. Upon relaxation, Bi atoms surrounding the divacancy form one octagonal ring and two adjacent pentagonal rings. Snapshots of divacancy taken at T=500 K indicate that Bi-Bi bonds between pentagonal and octagonal rings broken to form larger fourteen sided rings, while the overall character of the rest of honeycomb like structure is maintained. Spin-polarized calculations indicate that h-Bi with single vacancy or divacancy is nonmagnetic.

\begin{figure*}
\includegraphics[scale=0.8]{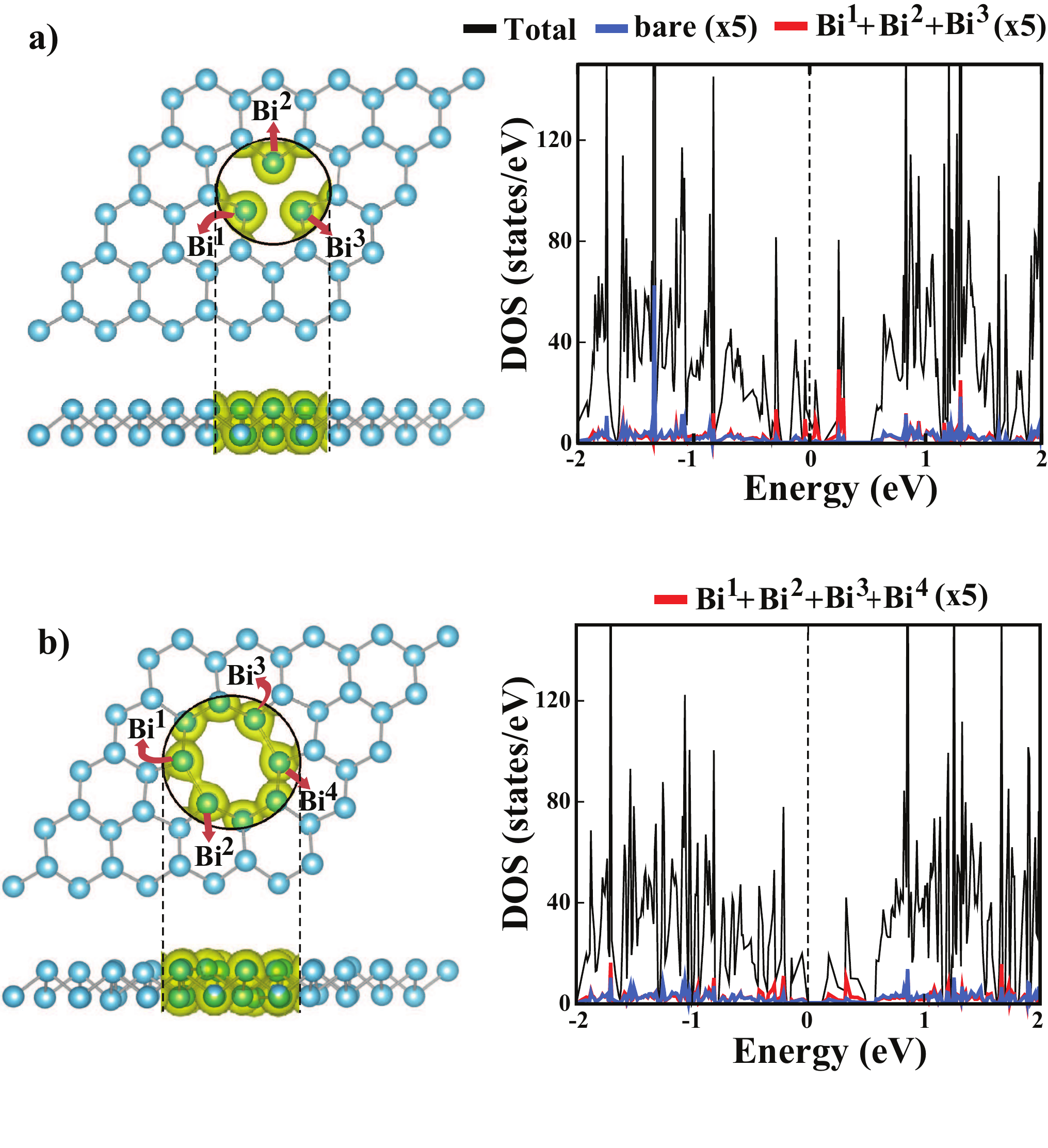}
\caption{(a) Optimized atomic configuration of a single vacancy in free standing SL h-Bi and charge density isosurfaces of surrounding three Bi atoms. The total (TDOS) densities of states of single vacancy+h-Bi, density of states projected to three Bi atoms surrounding the vacancy and density of states projected to a Bi atom farthest to the single vacancy representing extended and bare h-Bi. (b) Same for divacancy. The zero of energy is set at the Fermi level shown by vertical dashed line. The fundamental band gap of extended bare h-Bi is shaded.}
\label{fig7}
\end{figure*}

The charge density isosurfaces of the optimized atomic structure show that one dangling $sp^2$ orbitals oozes from surrounding atoms and constitutes chemically active sites. The analysis of calculated state densities in Fig.~\ref{fig7} indicates significant effects of the single vacancy on the electronic structure by the formation of states in the fundamental band gap and resonant states in the valence and conduction band continua near the band edges. These states induced by a single vacancy modify the electronic and optical properties by closing the band gap. In the case of divacancy in Fig.~\ref{fig7} (b), an octagonal ring with adjacent pentagonal rings are suitable to make all surrounding atoms three fold coordinated. Therefore, the divacancy of h-Bi has a larger hole as compared to a single vacancy, but chemically it is less active. When a divacancy formed, the localized gap states are expelled from the fundamental band gap, but they appear near the band edges due to saturation of dangling bonds.

\begin{figure*}
\includegraphics[scale=0.8]{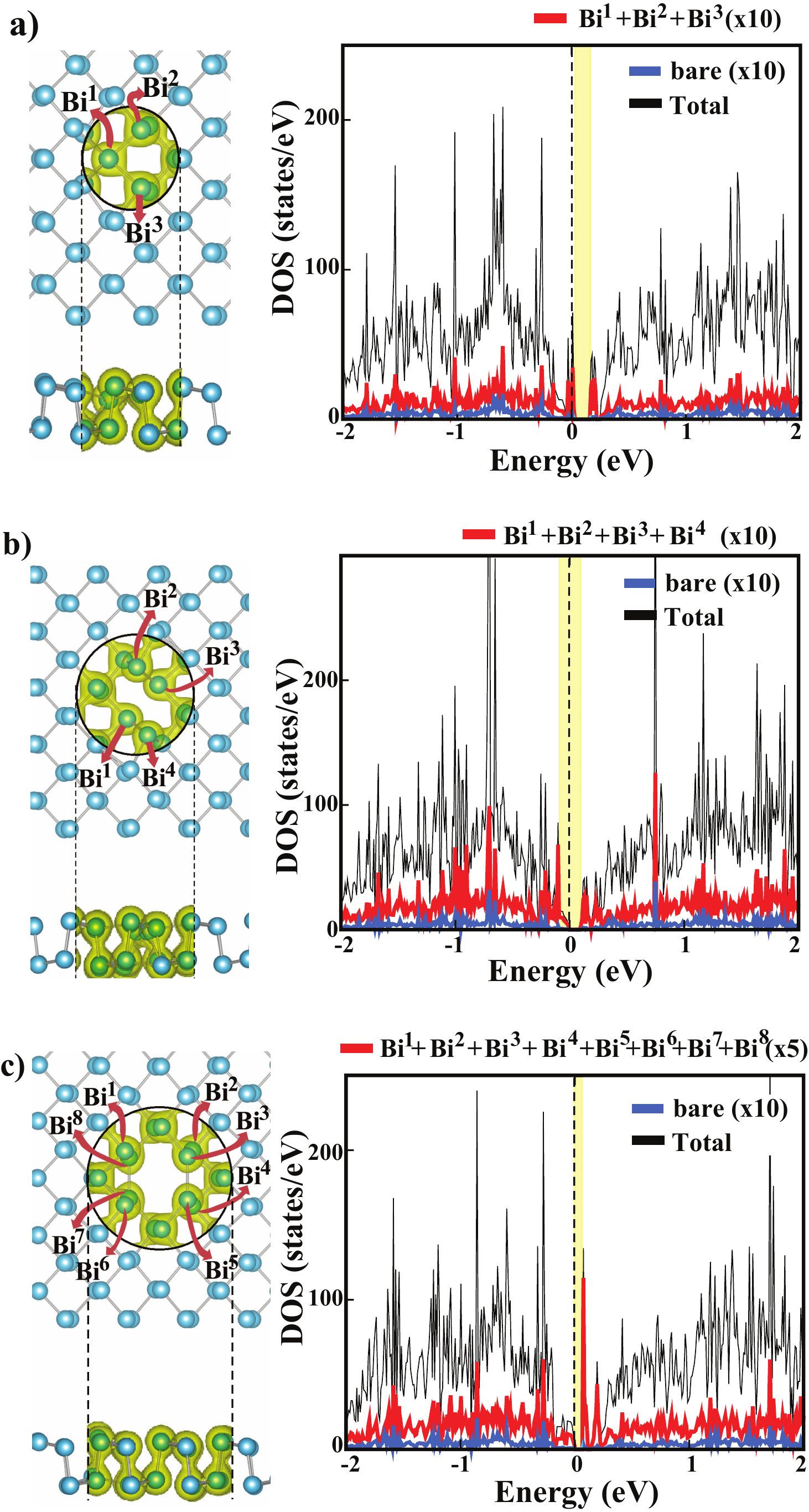}
\caption{(a) Optimized atomic configuration of a single vacancy in aw-Bi and charge density isosurfaces of surrounding three Bi atoms. The total (TDOS) densities of states of single vacancy+aw-Bi, density of states projected to three Bi atoms surrounding the vacancy and density of states projected to a Bi atom farthest to the single vacancy representing extended bare aw-Bi. (b) Same for a divacancy created in aw-Bi by removing a pair of nearest Bi atoms, which were forming a lateral bond at the top layer. (c) Same for a divacancy created in aw-Bi by removing a pair of Bi atoms, which were forming a vertical bond between top and bottom layer. The zero of energy is set at the Fermi level shown by vertical dashed line. The fundamental band gap of extended and bare aw-Bi is shaded.}
\label{fig8}
\end{figure*}

In SL aw-Bi two different types of single vacancies and divacancies can be created. In all these vacancies, the atomic structure collapses towards the center upon structure relaxation to form new bonds between Bi atoms, so that the lower coordination numbers around the vacancy are improved. Single vacancy can be formed at the top side either up-buckled or down-buckled atomic site with the formation energies in the range of $\sim$ 0.8 eV. Also, the divacancy can occur either by the removal of lateral nearest Bi atoms or perpendicular ones with the formation energies of 1.01 eV and 1.66 eV, respectively. In Fig.~\ref{fig8} (a) a single vacancy is created by removing the up-buckled atom. Single vacancy induces localized states in the fundamental band gap overlapping the valence and conduction band edges. This way the band gap of bare aw-Bi is slightly reduced. Upon creation of the divacancy by removing also nearest lateral Bi atom in Fig.~\ref{fig8} (b) localized states near the conduction band edge are expelled into the conduction band. The localized states induced by this lateral divacancy are located on Bi atoms surrounding the vacancy. If the divacancy is created by removing Bi atom below, the fundamental band gap is reduced by the localized states of the divacancy. These states are located at perpendicular Bi-Bi bonds surrounding the divacancy and constitute chemically active sites.

\section{Effect of point defects on band topology}
Further to the effects of adatoms and vacancies on the electronic and magnetic structures of SL bismuthenes, we next explore their effects on the topological phases of bare bismuthene. To this end, we consider only C adatom adsorbed to SL h-Bi and single vacancy and divacancy created in h-Bi, as example. The effects of both point defects on the electronic structure of bare  SL h-Bi were examined above by using the analysis of state densities with the premise that the coupling between nearest defects is weak to mimic single, isolated defect. Owing to the periodically repeating supercell method, this model also can be mapped to the band scheme. Under these circumstances, C adatom adsorbed to h-Bi is still a semiconductor with four valence electrons   forming flat bands. However, in the band picture, single vacancy in h-Bi is a metal forming bands crossing the Fermi level.

We calculated the Z$_2$ invariant corresponding to a single C adatom adsorbed to each 5$\times$5 supercell of h-Bi in order to reveal whether the non-trivial band topology of bare h-Bi will change upon adsorption. We found that the Z$_2$ invariant is 1, which means the band topology is not affected by a single C atom adsorbed to each 5$\times$5 supercell. Non-trivial band topology is protected by the time-reversal symmetry of the system.\cite{hasan} On the other hand, the non-trivial band topology of bare free standing SL h-Bi is changed  to trivial upon creating a single vacancy in each 5$\times$5 supercell, since the system is metallized. However, h-Bi having a divacancy in each 5$\times$5 regain non-trivial band topology with Z$_2$=1, since gap states are expelled upon rebondings around the divacancy.

\section{Discussions and Conclusions}
This study investigated the effects of point defects, namely adsorbed single adatoms, vacancy and divacancy on the physical properties of single-layer bismuthene structures. Adsorbed isolated adatoms we considered here readily form chemical bonds and give rise to localized and resonant states, which modify the electronic properties relevant for device application. When adatom-adatom coupling at high coverage becomes significant, these localized states can form bands in the band gap. The width of the impurity bands increases with their coupling; semiconducting bismuthene structures can even be metallized. Specific adatoms adsorbed along a row with suitable adatom-adatom distance can form strictly 1D metallic chains and 1D charge density. The dispersion and effective mass of these 1D electrons can be monitored by the adatom-adatom distance. At very low coverage, adsorbed adatoms function as if dopant with localized effects. Not only SL bismuthene, but also their bilayers, as well as multilayers can be functionalized by adatom adsorption.

Single vacancy as a point defect has small formation energy and hence can form readily at room temperature. Bismuth atoms surrounding the vacancy have $sp^2$ type dangling bonds oozing to the vacancy. Localized gap states and resonant states are derived from these dangling bonds and the electronic structure of the underlying bismuthene is locally modified. Additionally these dangling bonds make vacancy a chemically active site in bismuthene. Notably, the formation energy of divacancy is unexpectedly small owing to the rebondings of Bi atoms surrounding the divacancy. The most interesting outcome of rebondings is the removal of localized gap states as a result of the saturation of dangling bonds. Also in h-Bi, the divacancy transforms into Stone-Whales type defects consisting of two pentagonal and one octagonal rings. At high temperature, these defects transform into larger rings consisting of fourteen Bi atoms. When relaxed, interesting rings and defect states can also form through divacancies created in aw-Bi. In this respect, we believe that a study of patterned vacancies or holes in bismuthene can reveal important results.

While functionalization of 2D structures by point defects are commonly used to modify the electronic and magnetic structure, it is now of interest, whether the non-trivial band topology of a pristine 2D structure can be modified by the coverage of point defects. In this work, we found that pristine SL h-Bi is a non-trivial topological insulator. Then, whether and how this feature can be changed by the adsorption of selected adatoms and by the creation of single vacancies is of crucial importance. Of-course, one expects that single adatom or vacancy in a very large pristine SL h-Bi would not affect the existing topological phase. However, if the coverage of a point defect is high, localized states derived thereof can delocalized to form bands. The widths of bands are determined by the coupling of defects. This situation is closely related with the well-known Mott transition. In the present study, we found that the non-trivial band topology of pristine SL h-Bi is preserved after one C adatom adsorbed to each 5$\times$5 supercell. On the other hand, the non-trivial band topology disappeared upon the creation of a single vacancy in each supercell periodically, since h-Bi is metallized by the bands derived from the localized states of an isolated single vacancy. In contrast, the non-trivial band topology is maintained in the divacancy periodically repeated in the same supercell, since the bands derived from defect states removed from the band gap through rebondings of Bi atoms surrounding the divacancy.

In conclusion, we showed that not only optical and magnetic properties, but also topological features of pristine SL bismuthene can be modified by point defects (namely adatoms and vacancies). Modification of topological features depends on the energies of localized states in the fundamental band gap and also on the strength of coupling between point defects.

\begin{acknowledgments}
The computational resources are provided by TUBITAK ULAKBIM, High Performance and Grid Computing Center (TR-Grid e-Infrastructure). SC acknowledge financial support from the Academy of Sciences of Turkey (T\"{U}BA). This research was supported by Research Fund of the Adnan Menderes University under Project No.MF-16004. We thank Dominik Gresch for helpful discussions on operating Z2Pack.
\end{acknowledgments}

\end{document}